\documentclass{ws-procs9x6}
\newcommand{\eqb}{\begin{equation}}
\newcommand{\eqe}{\end{equation}}
\newcommand{\dmb}{\begin{displaymath}}
\newcommand{\dme}{\end{displaymath}}
\newcommand{\pd}{\partial}

\newcommand{\eab}{\begin{eqnarray}}
\newcommand{\eae}{\end{eqnarray}}
\newcommand{\ra}{\right\rangle}
\newcommand{\la}{\left\langle}

\newcommand{\be}{\begin{equation}}
\newcommand{\ee}{\end{equation}}

\newcommand{\La}{\Lambda}

\begin{document}

\title{Analytical approach to SU(2) Yang-Mills thermodynamics}

\author{R.~HOFMANN}

\address{Institut f\"ur Theoretische Physik, Universit\"at Heidelberg,  \\
Philosophenweg 16, \\ 
69120 Heidelberg, Germany, \\ 
E-mail: r.hofmann@thphys.uni-heidelberg.de}

\maketitle

\abstracts{We propose an analytical approach to SU(2) Yang-Mills 
thermodynamics. The existence of a macroscopic and rigid adjoint 
Higgs field, generated by dilute trivial-holonomy calorons at large 
temperature $T$ (electric phase), implies a twofold degeneracy 
of the ground state which signals a broken electric $Z_2$ symmetry. 
A finite energy density $\propto T$ of the ground state arises due to caloron interaction. 
An evolution equation for the effective gauge coupling, derived from thermodynamical self-consistency, 
predicts a second-order like transition (seen in lattice simulations) at $T_c$ 
to a phase where monopoles are 
condensed and off-Cartan excitations decoupled. In this magnetic phase the 
ground state is unique and dominates the pressure (negative total pressure). 
While the magnetic phase has a massive, propagating 
'photon' it confines fundamental matter (pre-confinement). 
The temperature dependence of the 
magnetic gauge coupling predicts the transition to 
the confining phase at $T_C\sim \frac{T_c}{1.9}$ where center-vortex loops condense 
and the 'photon' decouples. We believe that this transition 
is 'swallowed' by finite-size artefacts in lattice simulations. 
No thermodynamical connection exists between the 
confining and the magnetic phase.}

The objective is to propose a macroscopic, 
effective theory for SU(2) Yang-Mills thermodynamics which can be 
generalized to SU(N) \cite{Hofmann2004}. The approach is similar in spirit to the idea that 
superconductivity is macroscopically described by a U(1) Higgs theory 
\cite{GinzburgLandau1950,Abrikosov1957}. We predict 
the phase structure of SU(2) Yang-Mills theory, the 
(quasiparticle) spectrum of its excitations, and the $T$ dependence of 
thermodynamical quantities. As a result, the theory is shown to come in {\sl three} 
rather than two phases. We predict a vanishing entropy density 
and an equation of state $\rho=-P$ at $T_C$ and {\sl negative} pressure $P$ throughout 
the magnetic phase ($T_C\le T \le T_c$). An over-exponentially 
growing density of states (intersecting and single 
center-vortex loops) implies that the limiting temperature 
$T_C\sim \Lambda_{YM}$ of the 
confining phase can only then be exceeded if the spatial 
homogeneity of the system is sacrificed \cite{Hagedorn1965}. 
The situation of two disconnected 
thermodynamical regimes in a pure SU(2) gauge theory has an analogue 
in the ${\bf N}=1$ SUSY YM theory where a separating pole in the exactly known beta 
function exists \cite{NSVZ1986}.

\noindent The basic assumption is that a dilute-gas ensemble of calorons with 
trivial-holonomy (THC) \cite{HarringtonShepard1977} macroscopically forms a 
composite and 
adjoint Higgs field $\phi$ at high $T$ (electric phase). By dilute gas we mean 
that in a minimal, local definition (lowest possible mass dimension) of $\phi$,
\eqb
\label{locdefphi}
\phi^a(x)|\phi(x)|^5\equiv\la\mbox{tr}_{\tiny\mbox{N}} F_{\mu\nu}(x)\,t^a\,
F_{\nu\lambda}(x)F_{\lambda\mu}(x)\ra_{A_\beta^{\tiny\mbox{THC}}}\,,
\eqe
the average is performed over a single THC and its zero-mode 
deformations only. Fluctuations, that would lift the action of a 
given THC configuration $A_\beta^{\tiny\mbox{THC}}$ 
above the BPS bound, are discarded in Eq.\,(\ref{locdefphi}) because they 
mediate interactions between calorons to be considered later at the 
macroscopic level. The above assumption would be superfluous 
if it could be shown that at a given temperature $T$ 
nontrivial solutions to 
the gap-equation (\ref{locdefphi}) with $|\phi|=|\phi|(T)$ exist 
for a certain range of values of the fundamental gauge coupling $\bar{g}$ 
\footnote{The integration over the instanton scale $\bar{\rho}$ in Eq.\,(\ref{locdefphi}) 
is cut off in the UV at the compositeness scale $|\phi|^{-1}$.}. This is the objective of 
future research. While the (nonfluctuating, see below) 
field $\phi$ generates a $T$-dependent mass for topologically trivial 
modes $W^\pm$ on tree-level (thermal quasiparticles, solution to the 
infrared problem of thermal perturbation theory \cite{Linde1980}) the 'photon' 
$Z_0$ remains tree-level massless in the electric phase. 
The 'condensate' $|\phi|$ represents a compositeness scale governing 
the maximal off-shellness of gauge-modes 
and the maximal center-of-mass energy 
flowing into a vertex. This yields a convering loop expansion of thermodynamical 
quantities despite a large value of the effective coupling constant $e$.

A fixed color orientation of $\phi$ forms a finite domain induced by a 
`seed' caloron. At the points where at least four 
domains meet isolated zeros of $\phi$ exist with the associated magnetic monopoles being 
mappings from $S_2$ onto the coset spaces \{SU(2)/U(1)\} \cite{Kibble1976}. 
Microscopically, isolated monopoles are generated by 
nontrivial-holonomy calorons ($A_0(|\vec x|\to \infty)\not=0$ and regular gauge copies) 
which form and dissociate during 
domain collisions \cite{Nahm1984,KraanVanBaalNPB1998,GrossPisarskiYaffe1981,Diakonov2004,HoelbingRebbiRubakov2001}, see \cite{Korthals-Altes2001} for an estimate 
of the surface tension of the spatial Wilson loop in terms 
of screened magnetic monopoles. A fluctuation $A_\beta$ in the fundamental theory 
is decomposed into a BPS, topologically 
nontrivial part $A_\beta^{\tiny\mbox{THC}}$, building the ground-state, 
and topologically trivial $a_\beta$ associated with excitations: 
$A_\beta=A_\beta^{\tiny\mbox{THC}}+a_\beta$. The macroscopic action is
\eqb
\label{actE}
S_E=\int_0^{1/T}d\tau\hspace{-0.1cm}\int d^3x\,\left(\frac{1}{2}\,\mbox{tr}_{\tiny\mbox{N}}
\,G_{\mu\nu}G_{\mu\nu}+\mbox{tr}_{\tiny\mbox{N}}\,{D}_\mu\phi{D}_\mu\phi+
V_E(\phi)\right)\,,
\eqe
where $G^a_{\mu\nu}\equiv\pd_\mu a^a_\nu-\pd_\nu a^a_\mu-ef^{abc}a^b_\mu a^c_\nu$ 
denotes the field strength of a
top. trivial mode $a_\beta$, $e$ is the effective coupling constant, 
${D}_\beta\phi\equiv\pd_\beta+ie[\phi,a_\beta]$, and $\mbox{tr}_{\tiny\mbox{N}}\,t^a t^b\equiv 
1/2\,\delta^{ab}$. The potential 
$V_E(\phi)\equiv\mbox{tr}_{\tiny\mbox{N}} v_E^\dagger v_E$ is uniquely 
determined by the requirement of 
spatially homogeneous, periodic-in-euclidean-time $\tau$ solutions 
($\tau$ independent modulus, $0\le\tau\le 1/T$) 
to the BPS equation $\pd_{\tau}\phi=v_E$. Macroscopic BPS saturation derives 
from the zero-energy property of THCs (microscopically BPS) of which 
$\phi$ is composed (in absence of other gauge-field fluctuations), 
the other properties follow from thermodynamical equilibrium. The potential reads
\eqb
\label{potentialE}
V_E=\mbox{tr}\, v^\dagger_E v_E\equiv\Lambda_E^6\,
\mbox{tr}\,(\phi^2)^{-1}\,,
\eqe
where $\Lambda_E$ denotes a 
dynamically generated mass scale determined by a 
boundary condition to the thermodynamical evolution. Up to global gauge transformations 
the `square-root' $v_E$ is given as $v_E\equiv i\La_E^3 \lambda_1\phi/|\phi|^2$
where $\lambda_i\,(i=1,2,3)$ denote the Pauli matrices and $|\phi|\equiv 1/2\,\mbox{tr}\, \phi^2$. 
Solutions to the BPS 
equation $\pd_{\tau}\phi=v_E$ are labelled by nonzero winding 
numbers $l\in{\bf Z}$: 
\eqb
\label{solBPSE}
\hspace{-0.5cm}\phi_l(\tau)=\sqrt{\frac{\Lambda_E^3}{2\pi T |l|}}\,\lambda_3
\exp(-2\pi i T l\lambda_1\tau)\,.
\eqe
On the solutions (\ref{solBPSE}) we have $\pd^2_{|{\phi}_l|}V_E/T^2=12\pi^2\,l^2$ and 
$\pd^2_{|{\phi}_l|}V_E/|\phi_l|^2=3l^3\,\lambda_E^3\,$. 
We thus conclude that $\phi_l$ does not fluctuate thermodynamically and 
quantum mechanically ($\lambda_E\equiv 2\pi T/\La_E$ is much larger than unity, 
see \cite{Hofmann2004}). We restrict to lowest winding $|l|=1$. 
Interactions between THC are accounted for macroscopically 
by taking $\phi_1$ as a background in the equation of motion 
$D_\mu G_{\mu\beta}=2ie[\phi,{D}_\beta\phi_1]$. On the macroscopic level there must not be a 
net field strength in the thermal ground state, and thus the only admissible 
solution $a^{bg,1}_\beta$ is pure gauge, $G_{\mu\beta}[a^{bg,1}_\nu]=0=D_\beta\phi_1$. 
We have $a^{bg,1}_\beta=\frac{\pi}{e}\,T \delta_{\beta 4}\lambda_1$. On the configuration 
$a^{bg,1}_\beta$ the Polyakov loop is ${\bf P}=-{\bf 1}$. To asign ($T$ dependent 
and $\tau$ independent) masses 
$m_{W^\pm}^2=-2e^2\,\mbox{tr}\,[\phi,t^{1,2}][\phi,t^{1,2}]$ to off-Cartan 
fluctuations a {\sl singular} gauge transformation to unitary gauge needs to be 
performed \cite{Hofmann2004}. The Polyakov loop transforms as 
${\bf P}=-1\to {\bf P}=+1$. The singular gauge 
transformation does not change the {\sl periodicity} of the fluctuations 
$a_\beta$. Thus it is {\sl irrelevant} whether one integrates out $a_\beta$ 
in winding or unitary gauge in a loop expansion of thermodynamical 
quantities. The two distinct ground states ${\bf P}=\pm {\bf 1}\not=0$ together with the 
associated gauge-field fluctuations are 
physically equivalent. In unitary gauge there is a physical interpretation of 
$a_\beta$ and integrating them out is simple. The existence of two ground states 
signals a broken electric $Z_2$ symmetry and thus deconfinement. By virtue of 
$D_\beta\phi=0$ the vanishing energy density (pressure) of the hypothetical 
ground state, that is composed of noninteracting THCs only, is shifted to the 
finite energy density (pressure) (-)$V_E=4\pi\Lambda_E^3 T$ of the physical ground state 
by THC interaction. As a consequence, the covariant 
BPS equation $D_\tau\phi=v_E$ is not satisfied by the above 
configurations mirroring that ground-state physics on its own is thermodynamically 
incomplete (gauge-field {\sl excitations} are emitted and 
absorbed in addition). 

\noindent An expression for 
the total pressure $P$ in one-loop approximation is derived. Because of the compositeness 
scale $|\phi_1|$ quantum fluctuations can be neglected \cite{Hofmann2004}, and two-loop corrections are 
tiny ($<$2\% of the one-loop value \cite{HerbstHofmann2004}). Thermodynamics is consistent 
(as in the underlying theory) if we impose $\pd_a P=0$ where the mass parameter $a$ is defined as 
$a=2\pi e\lambda_E^{-3/2}$. This enforces an evolution of 
the effective gauge coupling with $T$ which, in implicit form, is governed by 
\eqb
\label{eeq}
\pd_a \lambda_E=-\frac{24\,\lambda_E^4\,a}{(2\pi)^6}D(2a)\,.
\eqe
The function $x D(x)$ denotes a scaled-out 
derivative w.r.t mass $T a$ of the thermal component of the one-loop pressure for a bosonic field. 
The right-hand side of Eq.\,(ref{eeq}) is zero at $a=0,\infty$ implying the existence of a highest and 
lowest attainable temperature. Shortly below $\lambda_{E,P}$ 
$e$ shows a large bump of height $\sim \lambda_{E,P}$ making the 
assumed importance of THCs in the partition function 
self-consistent due to their small action. A plateau 
value $e_{\tiny\mbox{Plateau}}=17.15$ (attractor) exists which is independent of the 
boundary condition imposed at large temperature $T_P$ (IR $\leftrightarrow$ UV decoupling, 
renormalizability of
the underlying theory). The constancy of $e$ after the formation of $\phi_1$ 
is consistent with the existence of isolated and locally conserved 
magnetic charge. We note here that the Stefan-Boltzmann limit is 
reached quickly (with three polarizations for the $W^\pm$ bosons!) 
due to a {\sl power}-like approach.

For lack of space we only quote the results for the magnetic 
phase (see \cite{Hofmann2004} for details) 
which sets in at $T_c$ where $a\to\infty$: condensation of magnetic monopoles; 
decoupling of $W^\pm$ bosons; 'photon' $Z_0$ is massless at $T_c$ and massive 
below; decoupling of $Z_0$, equation of state $\rho=-P$, and vanishing entropy density 
at $T_C\sim T_c/1.9$ (predicted from an evolution equation 
similar to Eq.\,(\ref{eeq}) for the magnetic coupling $g$); 
negative total pressure for $T_C\le T\le T_c$. The Polyakov loop 
is $+1$ already in winding gauge. Thus only a single ground state exists and 
the electric $Z_2$ symmetry is restored (confinement of fundamental test charges 
but existence of a massive, 
propagating 'photon' $Z_0$). In the confining phase, setting in at $T_C\sim \frac{T_c}{1.9}$ 
by violent 'reheating' (entropy generation), 
no propagating gauge mode exists, excitations are made of crossed and uncrossed closed magnetic 
flux lines (center-vortex loops), and the ground-state pressure is precisely zero 
once the vortex condensate has settled to the minimum of its potential. 
Since excitations resemble vacuum diagrams in a $\phi^4$ theory 
\cite{BenderWu1969} and since the mass spectrum of these states 
is equidistant it is easy to estimate that the density of states in 
the confining phase is over-exponentially growing. This implies the existence 
of the Hagedorn temperature $T_C$ and the thermodynamical 
disconnectedness of the confining phase from the other two phases.

\section*{Acknowledgments} 
The author would like to thank organizers and participants 
for a stimulating conference.

\end{document}